\documentclass[10pt]{article}
\usepackage{multicol}
\usepackage{graphicx}
\usepackage{amsmath}
\usepackage[a4paper]{geometry}
\usepackage{hyperref}
\usepackage{rotating}

\begin{document}

\begin{center}
{\Large \bf Freezeout properties of different light nuclei at the RHIC Beam Energy Scan}

\vskip1.0cm

M.~Waqas$^{1,}${\footnote{Corresponding author. Email (M.Waqas):
waqas\_phy313@yahoo.com; waqas\_phy313@ucas.ac.cn}},G. X. Peng$^{1,2,3}$ {\footnote{Corresponding author. Email (G. X. Peng): gxpeng@ucas.ac.cn}},
Rui-Qin Wang$^{4}${\footnote{E-mail: wangrq@qfnu.edu.cn}}, Muhammad Ajaz$^{5}${\footnote{E-mail: ajaz@awkum.edu.pk; muhammad.ajaz@cern.ch}}, Abd Al Karim Haj Ismail$^{6,7}${\footnote{E-mail: a.hajismail@ajman.ac.ae}},
\\

{\small\it  $^1$ School of Nuclear Science and Technology, University of Chinese Academy of Sciences,
Beijing 100049, Peoples Republic of China, 

$^2$ Theoretical Physics Center for Science Facilities, Institute of High Energy Physics, Beijing 100049, China,

$^3$ Synergetic Innovation Center for Quantum Effects \& Application, Hunan Normal University, Changsha 410081, China,

$^4$ Qufu Normal University, Qufu 273165, Shandong, Peoples Republic of China,

$^5$ Department of Physics, Abdul Wali Khan University Mardan, 23200 Mardan, Pakistan,

$^6$ Department of Mathematics and Science, Ajman University, PO Box 346, UAE,

$^7$ Nonlinear Dynamics Research Center (NDRC), Ajman University, PO Box 346, UAE}
\end{center}

\vskip1.0cm

{\bf Abstract:} We study the transverse momentum spectra of light nuclei (deuteron, anti-deuteron and triton) produced in Gold-Gold (Au-Au) collisions in different centrality bins by the blast wave model with Tsallis statistics. The model results are in agreement with the experimental data measured by STAR Collaboration in special transverse momentum ranges. We extracted the kinetic freezeout temperature, transverse flow velocity and kinetic freezeout volume. It is observed that kinetic freezeout temperature and transverse flow velocity increases initially, and then saturates from 14.5~--~39 GeV, while the kinetic freezeout volume increase initially up to 19.6 GeV but saturates from 19.6~--~39 GeV. This may indicate that the phase transition starts in part volume that ends in the whole volume at 39 GeV and the critical point may exists somewhere in the energy range of 14.5~--~39 GeV. The present work observed that the kinetic freezeout temperature, transverse flow velocity and kinetic freezeout volume has a decreasing trend from central to peripheral collisions. We found the freezeout volume of triton is smaller than those of deuteron and anti-deuteron, which shows that triton freezeout earlier than that of deuteron and anti-deuteron.
\\



\vskip1.0cm

\begin{multicols}{2}

{\section{Introduction}}
The identification of various phases of dense matter is one of the most important problem in high energy
collisions. The quark-gluon plasma (QGP) is a droplet of deconfined matter formed at very high temperature and
density in high energy collisions. A similar phase transition was occured in the early universe shortly
after the big-bang prior to condensation of hadrons. The creation of QGP and to study its properties is the
main aim of relativistic heavy ion collisions. The collisions at Beam Energy Scan (BES) program at the
Relativistic Heavy Ion Collider (RHIC) offers a unique possibility to explore the quantum chromodynamics
(QCD) phase diagram \cite{1,2,3,4}. The usual phase diagram of QCD is plotted as chemical freezeout temperature ($T_{ch}$) against
the baryon chemical potential ($\mu_B$). Consider a thermalized system created in heavy ion collisions, the
change in $T_{ch}$ and $\mu_B$ is expected with varying the collision energy \cite{5,6,7}. It is suggested by several
QCD-based models \cite{8,9,10,11} and the calculations from the lattice QCD \cite{9} that if the system created in collisions
correspond to larger values for $\mu_B$, first order phase transition will occur. The point in the QCD
phase diagram plane where the first order phase transition ends is said to be the QCD critical point \cite{12,13}.
Besides, the kinetic freezeout volume is an important parameter to be measured. Consider a certain volume contains baryon.
It is a common sense that baryons have non-zero spatial volume
\cite{14}. This clearly suggests the existence of critical volume, where the baryons fill the volume completely. It
is supposed that the baryon structure disappear at the critical volume, and forms QGP.

Since beginning, the exploration of the production of light nuclei has been an active part in the studies of heavy ion
collisions. The collided nuclei at relativistic heavy ion collisions are mainly creating a zone of large energy
density and temperature. This is nearly baryon number free \cite{15}.

The underlying process of production of light nuclei is not well understood \cite{16,17,18}. Fortunately, some phenomenological models have proved to be particularly successful in describing the production of light nuclei. One group gives its approach through coalescence of produced or participant nucleons and describe some observable such as $p_T$ spectra and flows of light nuclei very well \cite{19,20,21,22,23,1a,2a,3a,4a,5a,6a,7a}. The other is statistical hadronization models, which successfully reproduce the measured yields as well as yield ratios of different light nuclei \cite{24,25,8a,9a,10a,11a,12a,13a}. Dynamical models based on the kinetic theory have also been used to describe different production characteristics of light nuclei \cite{26,14a,15a,16a,17a,18a}.
Since the binding energy of light nuclei is small, they can not survive when the temperature is much higher than their binding energy. For light hadrons the typical kinetic freezeout temperature ($T_0$) is around 100 MeV  \cite{16}, therefore, they might disintegrate
and be formed again by the final state coalescence the nucleons are de-coupled from the system. Hence the production of these light nuclei can be used for the extraction of important information of nucleons distribution at freezeout \cite{20,23,27}. In addition, a comprehensive study on the subject of the light (anti-)nuclei production in hh and AA collisions at LHC energies are given in \cite{28} where the authors concluded that the measured production yields of light anti-hyper nuclei clearly match the expectations of the thermal model described in \cite{29}.

The study of transverse momentum spectra of the particles is very important because it gives some useful information
about the temperature (effective temperature, initial temperature and kinetic freezeout temperature), kinetic freezeout
volume ($V$) and transverse flow velocity ($\beta_T$) of the final state particles. The energy and centrality
dependence of kinetic freezeout temperature and transverse flow velocity in high energy collisions has been an
interesting subject in the community. Multiple studies agreed in the increasing trend of $T_0$ and $\beta_T$ with
increasing the collision energy. For $\beta_T$ most studies show an increasing trend with collision energy \cite{30,31,32,33,34,35,36}.
For $T_0$, several studies declare its increasing trend with increasing the collision energy \cite{32,33}, while others
argue its decreasing trend \cite{34,36,37,38,39}  with energy and centrality \cite{39a} and some affirm a little bit dependence on collisions energy \cite{30,31}. Similarly
the dependence of $T_0$ and $\beta_T$ on centrality has also a complex structure. Several studies claimed the decrease
of $T_0$ \cite{32,40,41,42} with decreasing centrality, while others stated an increasing trend of $T_0$ \cite{36,43} with decreasing the
centrality. For $\beta_T$, most of the studies agree to decrease with decreasing centrality \cite{36,40,41,42,43,44,45,46}. Besides, the
study of kinetic freezeout volume is important. Most of studies agreed on larger $V$ in central collisions and at higher
energies which decrease with decreasing centrality and collision energy \cite{47,48,49}.

In this paper, we will analyse light nuclei (Deuteron and Triton) and will extract the kinetic freezeout temperature,
kinetic freezeout volume and transverse flow velocity by using the blast wave model with Tsallis statistics.
\\

{\section{The method and formalism}}
We used the blast wave model with Tsallis statistics (TBW) which is modified from the original blast wave model with
Boltzmann Gibbs statistics (BGBW) when a Tsallis statistics displaces the standard Boltzmann Gibbs statistics
for the particle ejection distribution.

The blast wave model with Tsallis statistics \cite{50,51,52,53} results in the $p_T$ distribution to be
\begin{align}
f_S(p_T)=&\frac{1}{N}\frac{\mathrm{d}N}{\mathrm{d}p_\mathrm{T}}= \frac{1}{N}\frac{gV}{(2\pi)^2} p_T m_T \int_{-\pi}^\pi d\phi\int_0^R rdr \nonumber\\
& \times\bigg\{{1+\frac{q-1}{T_0}} \bigg[m_T  \cosh(\rho)-p_T \sinh(\rho) \nonumber\\
& \times\cos(\phi)\bigg]\bigg\}^\frac{-q}{(q-1)}
\end{align}
where N represents the number of particles, $g$ is the degeneracy factor, $V$ is the kinetic freeze out volume,
$m_T$ is the transverse mass ($m_T=\sqrt{p_T^2+m_0^2}$), $m_0$ is the rest mass of the particle. $\phi$ denotes the
particle emission angle in the rest frame of thermal source, $q$ denotes entropy that characterize the degree of
non-equilibrium of the produced system which is a new parameterization suggested in TBW compared to BGBW model. $r$
is the radial coordinates and $R$ denotes maximum $r$. $\rho=\tanh^{-1} [\beta(r)]$ is the boost angle, $\beta(r)=\beta_S(r/R)^{n_0}$
is a self-similar flow profile, where $\beta_S$ is the flow velocity on the surface, as mean $\beta_r$,
$\beta_T=(2/R^2)\int_0^R r\beta(r)dr=2\beta_S/(n_0+2)=2\beta_S/3$, and $n_0$ is taken to be 1 \cite{30}. In eqs. (1), the index $q/(q-1)$
\cite{54} is the replacement of $1/(q-1)$ used in Ref. \cite{50,55} because $q$ is also required for the thermodynamics
consistency \cite{54,56}. This displacement causes a very little difference in the values of $q$ in the two cases because $q$ is
being close to 1.  Furthermore, another parameter $N_0$ has been defined in table 1 in the next section. In fact $N_0$ is only a normalization factor (which is used to compare the fit function $f_S(p_T)$ and the experimental spectra) and the data are not cross-section, but they are proportional to the volumes sources of producing various particles. Therefore, it is significant to study $N_0$.

 It is noteworthy that, there are various distributions which can be used to describe the particle spectra i.e Blast wave model with Boltzmann Gibbs statistics, blast wave model with statistics, Tsallis statistics, Standard distribution and Hagedorn thermal model etc. Among them the Tsallis distribution and its
alternative forms can fit the wider spectra. Particularly, the Tsallis distribution can cover the multi-component
(two or three-component) standard distribution \cite{57}. This means that, to fit the spectra as widely as possible,
one can use the Tsallis distribution and its alternative forms. In addition, the blast-wave fit can be easily used
to extract synchronously the bulk properties in terms of kinetic freeze-out temperature $T_0$, kinetic freezeout volume $V$
and transverse flow velocity $\beta_T$ . Therefore, it is suitable for us to use the combination of the Tsallis
distribution and the blast-wave fit to extract $T_0$, $V$  and $\beta_T$ from wide enough spectra. This combination is
in deed the blast-wave fit with Tsallis statistics \cite{50,58}.
\\

{\section{Results and discussion}}
The transverse momentum spectra $p_T$ spectra, $1/N_{ev}$[(1/2$\pi$ $p_T$) $d^2$ $N$/$dy$ $dp_T$] of deuteron produced at collision different energies in Au-Au collisions are shown in figure 1. The symbols are used for the representation of the experimental data measured by STAR Collaboration \cite{59} in the mid-rapidity range of $|y|<0.3$ and the curves are our fitting results by using the blast-wave model with Tsallis statistics, Eq. (1). The solid curve represents individual fitting and the dashed curves in some spectra show the global fitting. The spectra is distributed in various centrality classes, and some of them are re-scaled which are mentioned in figure. The corresponding ratio of data/fit is followed in each panel. The ratio of data/fit obtained from the individual fit are demonstrated by filled symbols while the empty symbols represent the data/fit ratio obtained from the global fitting. The results of data/fit for the individual (global) fit are presented in the panels to monitor the departure of the individual (global) fit from data. The method of least squares is used in each fitting in a special $p_T$ range to obtain the best values of parameters. The extracted parameters along with $\chi^2$ and degree of freedom (dof) are listed in Table 1. It can be seen that Eq. (1) fits well the data in Au-Au collisions at 7.7, 11.5, 14.5, 19.6, 27, 39, 62.4 and 200 GeV.
\begin{figure*}[htbp]
\begin{center}
\includegraphics[width=14.cm]{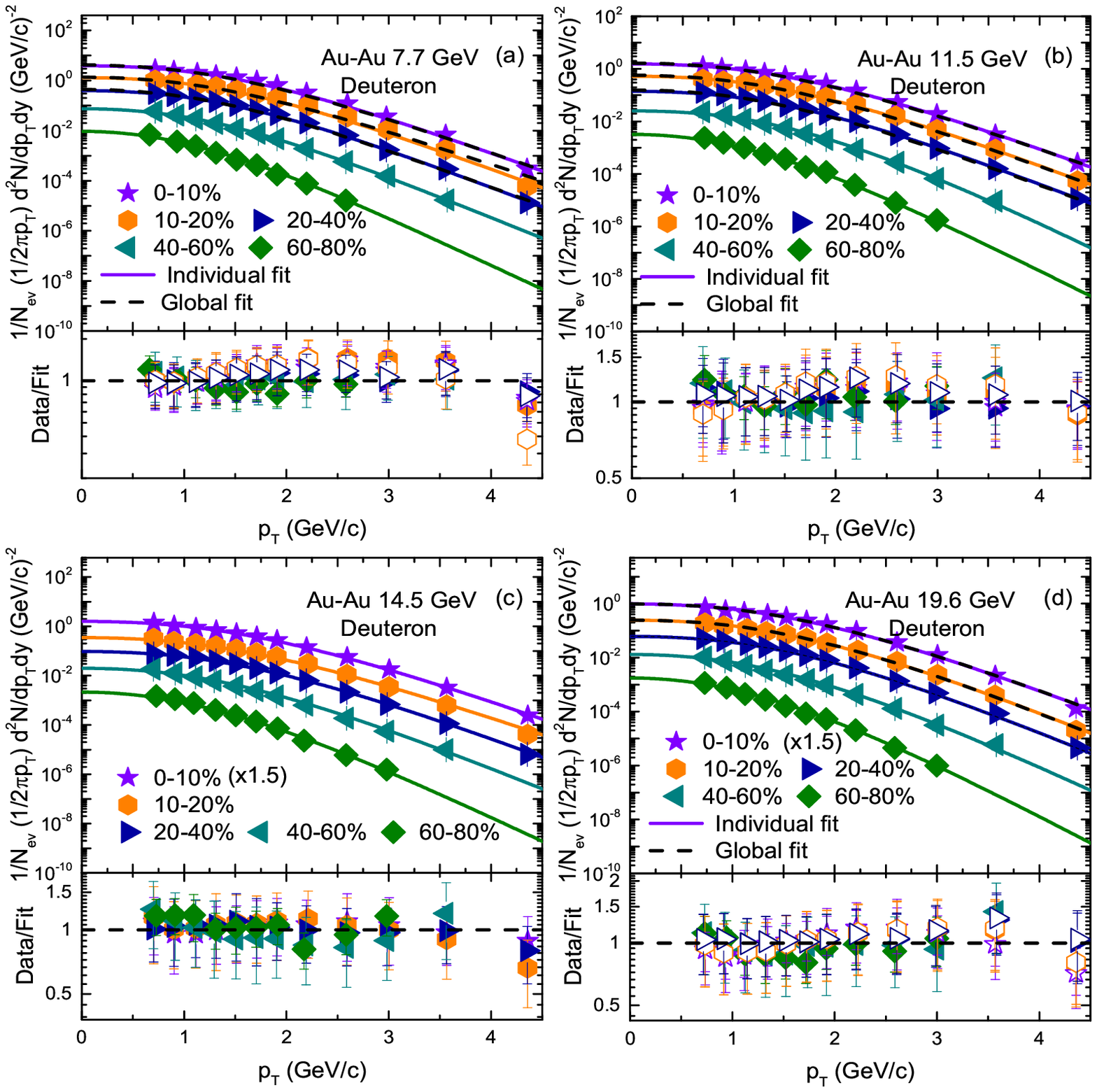}
\end{center}
continue
\end{figure*}
\begin{figure*}[htbp]
\begin{center}
\includegraphics[width=14.cm]{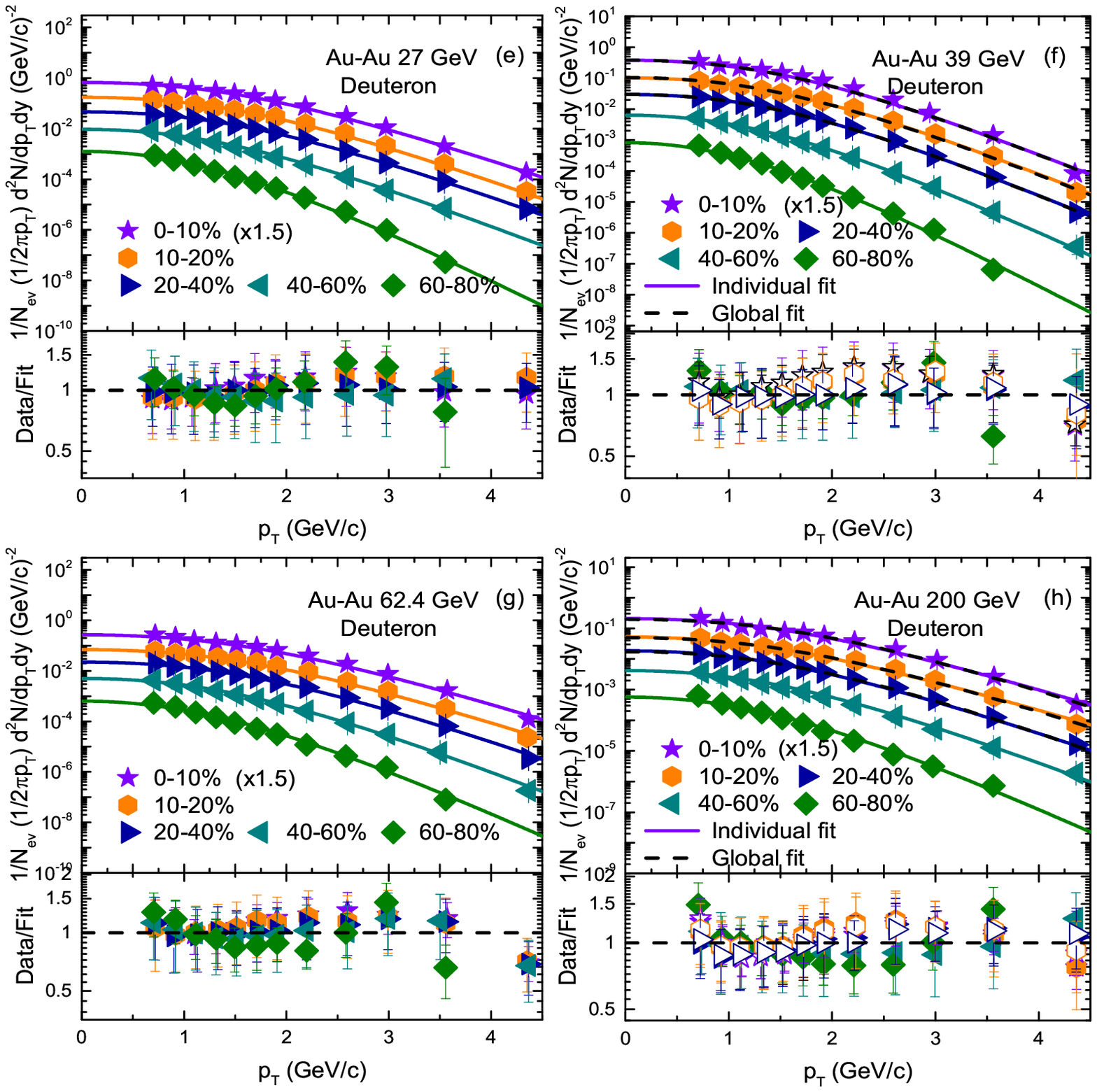}
\end{center}
Fig.1. Transverse momentum spectra of deuteron produced at different energies in different centrality intervals in Au-Au, where $1/N_{ev}$ on the vertical axis denote the number of events. The symbols demonstrate the experimental data of STAR Collaboration measured in mid-rapidity $|y|<0.3$ \cite{59}. The curves are our results calculated by using the blast wave model with Tsallis statistics. The solid and dashed curves represents the individual and global fitting respectively.
\end{figure*}
\begin{table*}
{\scriptsize Table 1. Values of free parameters ($T_0$ and
$\beta_T$), entropy index (q), normalization constant ($N_0$),
$\chi^2$, and degree of freedom (dof) corresponding to the curves
in Figs. 1, 2 and 3. \vspace{-.50cm}
\begin{center}
\begin{tabular}{ccccccccccc}\\ \hline\hline
Collisions       & Centrality   & $T_0$ (GeV)       & $\beta_T$ (c)       & $V (fm^3)$      & $q$              &  $N_0$                       & $\chi^2$/ dof \\ \hline
Fig. 1           & 0--10\%      &$0.112\pm0.006$  & $0.352\pm0.009$   & $3400\pm165$    & $1.014\pm0.004$  &$0.042\pm0.004$                     & 5/7\\
Au-Au            & 10--20\%     &$0.108\pm0.007$  & $0.342\pm0.011$  & $3200\pm150$    & $1.019\pm0.003$  &$0.014\pm0.005$                      & 10/7\\
7.7 GeV          & 20--40\%     &$0.104\pm0.005$  & $0.325\pm0.010$  & $3000\pm128$    & $1.016\pm0.005$  &$0.0043\pm0.0006$                    & 2/7\\
Deuteron         & 40--60\%     &$0.101\pm0.006$  & $0.290\pm0.009$  & $2795\pm162$    & $1.015\pm0.004$  &$7\times10^{-4}\pm4\times10^{-5}$    & 0.5/6\\
                 & 60--80\%     &$0.096\pm0.006$  & $0.260\pm0.007$  & $2550\pm171$    & $1.01\pm0.005$  &$6.8\times10^{-5}\pm6\times10^{-6}$  & 3/4\\
\cline{2-8}
 Au-Au           & 0--10\%     &$0.125\pm0.005$  & $0.365\pm0.006$  & $4000\pm170$     & $1.01\pm0.005$  &$0.0155\pm0.004$                       & 1.4/7\\
 11.5 GeV        & 10--20\%    &$0.120\pm0.006$  & $0.353\pm0.007$  & $3830\pm185$     & $1.011\pm0.002$ &$0.0052\pm0.0005$                     & 2/7\\
                 & 20--40\%    &$0.116\pm0.007$  & $0.340\pm0.009$  & $3615\pm180$     & $1.013\pm0.003$ &$0.0014\pm0.0004$                   & 1/7\\
                 & 40--60\%    &$0.111\pm0.005$  & $0.310\pm0.008$  & $3400\pm178$     & $1.01\pm0.003$ &$2.3\times10^{-4}\pm3\times10^{-5}$    & 0.7/6\\
                 & 60--80\%    &$0.107\pm0.007$ & $0.275\pm0.010$   & $3200\pm180$     & $1.005\pm0.004$ &$2\times10^{-5}\pm4\times10^{-6}$     & 2/5\\
\cline{2-8}
  Au-Au          & 0--10\%    &$0.137\pm0.005$  & $0.380\pm0.010$  & $4700\pm250$     & $1.0005\pm0.0003$&$0.0091\pm0.0005$                     & 2.6/7\\
  14.5 GeV      & 10--20\%    &$0.130\pm0.006$  & $0.372\pm0.009$  & $4500\pm210$      & $1.006\pm0.004$  &$0.003\pm0.0004$                     & 2.6/7\\
                 & 20--40\%   &$0.127\pm0.006$  & $0.365\pm0.006$  & $4320\pm200$      & $1.003\pm0.005$  &$8\times10^{-4}\pm4\times10^{-5}$    & 1/7\\
                 & 40--60\%   &$0.124\pm0.004$  & $0.327\pm0.009$  & $4140\pm180$      & $1.004\pm0.003$  &$1.4\times10^{-4}\pm5\times10^{-5}$  & 1.5/6\\
                 & 60--80\%   &$0.120\pm0.007$  & $0.290\pm0.011$  & $4000\pm190$      & $1.001\pm0.000$  &$1.1\times10^{-5}\pm5\times10^{-6}$  & 12/5\\
\cline{2-8}
  Au-Au          & 0--10\%    &$0.138\pm0.007$  & $0.380\pm0.009$  & $5270\pm230$     & $1.0001\pm0.0003$  &$0.005\pm0.0004$                      & 2.7/7\\
  19.6 GeV       & 10--20\%   &$0.131\pm0.008$  & $0.373\pm0.011$  & $5136\pm200$     & $1.001\pm0.004$    &$0.0018\pm0.0005$                     & 1.2/7\\
                 & 20--40\%   &$0.128\pm0.007$  & $0.365\pm0.010$  & $4900\pm210$     & $1.002\pm0.003$    &$4.5\times10^{-4}\pm4\times10^{-5} $  & 0.8/7\\
                 & 40--60\%   &$0.125\pm0.006$  & $0.326\pm0.011$  & $4700\pm190$     & $1.002\pm0.003$    &$8\times10^{-5}\pm5\times10^{-6}$     & 1/6\\
                 & 60--80\%   &$0.121\pm0.005$  & $0.290\pm0.007$  & $4500\pm180$     & $1.003\pm0.005$    &$7.8\times10^{-6}\pm4\times10^{-7}$   & 12/5\\
\cline{2-8}
  Au-Au         & 0--10\%    &$0.138\pm0.007$  & $0.381\pm0.010$   & $5274\pm200$     & $1.004\pm0.004$     &$0.0035\pm0.0004$                      & 1.6/7\\
  27 GeV        & 10--20\%   &$0.131\pm0.005$  & $0.373\pm0.012$   & $5140\pm210$     & $1.004\pm0.005$     &$0.0013\pm0.0005$                      & 1/7\\
                 & 20--40\%   &$0.128\pm0.006$ & $0.366\pm0.010$   & $4906\pm215$     & $1.006\pm0.004$     &$3.5\times10^{-4}\pm4\times10^{-5}$    & 0.2/7\\
                 & 40--60\%   &$0.125\pm0.009$ & $0.326\pm0.009$   & $4700\pm180$     & $1.008\pm0.005$     &$6\times10^{-5}\pm5\times10^{-6}$      & 0.7/6\\
                 & 60--80\%   &$0.121\pm0.008$ & $0.290\pm0.008$  & $4500\pm190$     & $1.0001\pm0.0002$    &$5.8\times10^{-6}\pm5\times10^{-7}$    & 6.7/6\\
\cline{2-8}
  Au-Au         & 0--10\%   &$0.138\pm0.006$  & $0.381\pm0.010$   & $5273\pm206$     & $1.005\pm0.005$      &$0.002\pm0.0004$                      & 5.6/7\\
  39 GeV        & 10--20\%  &$0.131\pm0.005$  & $0.370\pm0.011$   & $5140\pm203$     & $1.009\pm0.004$      &$8\times10^{-4}\pm5\times10^{-5}$     & 2/7\\
                 & 20--40\%  &$0.128\pm0.006$ & $0.366\pm0.009$   & $4904\pm205$     & $1.0085\pm0.0005$     &$2.3\times10^{-4}\pm4\times10^{-5}$   & 0.5/7\\
                 & 40--60\%  &$0.125\pm0.007$ & $0.327\pm0.010$   & $4702\pm205$     & $1.0085\pm0.0005$     &$4.1\times10^{-5}\pm5\times10^{-6}$   & 0.3/7\\
                 & 60--80\%  &$0.121\pm0.008$ & $0.290\pm0.008$   & $4500\pm200$     & $1.009\pm0.003$      &$4\times10^{-6}\pm4\times10^{-7}$     & 5.7/6\\
\cline{2-8}
  Au-Au         & 0--10\%  &$0.154\pm0.005$   & $0.400\pm0.007$  & $6000\pm220$    & $1.002\pm0.003$    &$0.0014\pm0.0004$                         & 4/7\\
  62.4 GeV       & 10--20\% &$0.148\pm0.007$  & $0.390\pm0.008$  & $5800\pm200$    & $1.002\pm0.004$    &$5.3\times10^{-4}\pm6\times10^{-5}$       & 2.7/7\\
                 & 20--40\% &$0.143\pm0.006$  & $0.380\pm0.010$  & $5600\pm200$    & $1.001\pm0.002$    &$1.6\times10^{-4}\pm5\times10^{-5}$       & 3/7\\
                 & 40--60\% &$0.139\pm0.007$  & $0.350\pm0.009$  & $5400\pm215$    & $1.0001\pm0.0003$  &$3\times10^{-5}\pm4\times10^{-6}$         & 3.4/7\\
                 & 60--80\% &$0.135\pm0.005$  & $0.310\pm0.011$  & $5187\pm210$    & $1.001\pm0.004$    &$3\times10^{-6}\pm5\times10^{-7}$         & 8/6\\
\cline{2-8}
  Au-Au         & 0--10\%   &$0.160\pm0.006$  & $0.430\pm0.009$  & $6800\pm180$    & $1.002\pm0.003$     &$0.0011\pm0.0004$                         & 5.7/7\\
  200 GeV       & 10--20\%  &$0.155\pm0.007$  & $0.415\pm0.010$  & $6600\pm205$    & $1.008\pm0.004$    &$4\times10^{-4}\pm4\times10^{-5}$          & 1.2/7\\
                 & 20--40\% &$0.151\pm0.005$  & $0.401\pm0.011$  & $6410\pm190$    & $1.006\pm0.004$    &$1.3\times10^{-4}\pm5\times10^{-5} $       & 0.7/7\\
                 & 40--60\% &$0.145\pm0.006$  & $0.368\pm0.009$  & $6230\pm215$    & $1.007\pm0.003$    &$2.7\times10^{-5}\pm5\times10^{-6}$        & 1.3/7\\
                 & 60--80\% &$0.140\pm0.005$  & $0.334\pm0.010$  & $6000\pm200$    & $1.004\pm0.003$    &$3\times10^{-6}\pm3\times10^{-7}$         & 8.5/6\\
\hline
Fig. 2           & 0--10\%     &$0.118\pm0.007$  & $0.365\pm0.010$  & $4300\pm180$     & $1.024\pm0.005$  &$6\times10^{-6}\pm5\times10^{-7}$     & 1/0\\
 Au-Au           & 10--20\%    &$0.114\pm0.006$  & $0.353\pm0.011$  & $4150\pm190$     & $1.001\pm0.004$  &$2.7\times10^{-6}\pm4\times10^{-7}$   & 13/0\\
 11.5 GeV        & 20--40\%    &$0.111\pm0.006$  & $0.340\pm0.010$  & $4000\pm200$     & $1.0002\pm0.005$ &$5.7\times10^{-7}\pm5\times10^{-8}$   & 3/0\\
 Anti-deuteron   & 40--60\%    &$0.107\pm0.007$  & $0.309\pm0.011$  & $3800\pm185$     & $1.002\pm0.004$  &$3\times10^{-7}\pm4\times10^{-8}$    & 0.7/0\\
\cline{2-8}
  Au-Au          & 0--10\%    &$0.130\pm0.007$  & $0.380\pm0.012$  & $4940\pm200$      & $1.07\pm0.005 $  &$1.4\times10^{-5}\pm6\times10^{-6}$  & 2/1\\
  14.5 GeV       & 10--20\%   &$0.126\pm0.005$  & $0.372\pm0.008$  & $4800\pm222$      & $1.001\pm0.006$  &$4.2\times10^{-6}\pm7\times10^{-7}$  & 5/1\\
                 & 20--40\%   &$0.120\pm0.007$  & $0.365\pm0.010$  & $4600\pm188$      & $1.01\pm0.004$   &$2.2\times10^{-6}\pm6\times10^{-7}$  & 1/1\\
                 & 40--60\%   &$0.117\pm0.006$  & $0.328\pm0.011$  & $4400\pm193$      & $1.005\pm0.005$  &$7\times10^{-7}\pm5\times10^{-8}$    & 1/0\\
                 & 60--80\%   &$0.113\pm0.005$  & $0.290\pm0.011$  & $4200\pm205$      & $1.001\pm0.005$  &$9\times10^{-8}\pm6\times10^{-9}$    & 5/0\\
\cline{2-8}
  Au-Au          & 0--10\%    &$0.130\pm0.006$  & $0.380\pm0.010$  & $5600\pm210$     & $1.005\pm0.006$    &$3.2\times10^{-5}\pm5\times10^{-6}$   & 18/5\\
  19.6 GeV       & 10--20\%   &$0.126\pm0.005$  & $0.373\pm0.009$  & $5400\pm187$     & $1.008\pm0.004$    &$1.2\times10^{-5}\pm4\times10^{-6}$   & 1.2/4\\
                 & 20--40\%   &$0.120\pm0.006$  & $0.365\pm0.012$  & $5180\pm180$     & $1.005\pm0.005$    &$1.5\times10^{-6}\pm5\times10^{-6} $   & 2/4\\
                 & 40--60\%   &$0.118\pm0.007$  & $0.326\pm0.010$  & $5000\pm200$     & $1.005\pm0.004$    &$1.2\times10^{-6}\pm4\times10^{-7}$   & 2/4\\
                 & 60--80\%   &$0.114\pm0.006$  & $0.290\pm0.010$  & $4800\pm200$     & $1.001\pm0.005$    &$2\times10^{-7}\pm5\times10^{-5}$     & 12/5\\
\cline{2-8}
  Au-Au         & 0--10\%    &$0.130\pm0.007$  & $0.378\pm0.011$   & $5600\pm225$     & $1.001\pm0.007$     &$7.5\times10^{-5}\pm4\times10^{-6}$   & 2/6\\
  27 GeV        & 10--20\%   &$0.126\pm0.005$  & $0.373\pm0.012$   & $5400\pm200$     & $1.008\pm0.007$     &$3\times10^{-5}\pm6\times10^{-6}$     & 2.5/6\\
                 & 20--40\%  &$0.120\pm0.005$  & $0.365\pm0.013$   & $5180\pm208$     & $1.009\pm0.006$     &$1.2\times10^{-5}\pm5\times10^{-6}$   & 1/6\\
                 & 40--60\%  &$0.118\pm0.005$ & $0.326\pm0.011$    & $5000\pm200$     & $1.007\pm0.004$     &$2.8\times10^{-6}\pm6\times10^{-7}$   & 4/5\\
                 & 60--80\%  &$0.114\pm0.005$ & $0.290\pm0.009$    & $4800\pm170$     & $1.001\pm0.004$     &$4\times10^{-7}\pm5\times10^{-8}$     & 3/3\\
\cline{2-8}
  Au-Au         & 0--10\%   &$0.130\pm0.007$  & $0.380\pm0.012$   & $5600\pm220$     & $1.015\pm0.005$      &$1.35\times10^{-4}\pm3\times10^{-6}$  & 5.6/7\\
  39 GeV        & 10--20\%  &$0.126\pm0.006$  & $0.373\pm0.010$   & $5400\pm210$     & $1.011\pm0.006$      &$6\times10^{-5}\pm6\times10^{-6}$     & 2/6\\
                 & 20--40\%  &$0.120\pm0.006$ & $0.365\pm0.012$   & $5180\pm225$     & $1.01\pm0.006$      &$2\times10^{-5}\pm3\times10^{-6}$     & 1/6\\
                 & 40--60\%  &$0.118\pm0.005$ & $0.327\pm0.012$   & $5000\pm213$     & $1.006\pm0.005$      &$5\times10^{-6}\pm6\times10^{-7}$     & 0.5/6\\
                 & 60--80\%  &$0.114\pm0.006$ & $0.290\pm0.010$   & $4800\pm200$     & $1.001\pm0.004$      &$6.8\times10^{-7}\pm5\times10^{-8}$   & 5/4\\
\cline{2-8}
  Au-Au         & 0--10\%  &$0.142\pm0.008$   & $0.399\pm0.009$  & $6300\pm229$    & $1.01\pm0.006$      &$2.3\times10^{-4}\pm4\times10^{-5}$      & 5/7\\
  62.4 GeV       & 10--20\% &$0.138\pm0.005$  & $0.388\pm0.011$  & $6100\pm210$    & $1.014\pm0.006$     &$9\times10^{-5}\pm3\times10^{-6}$        & 1.6/6\\
                 & 20--40\% &$0.134\pm0.006$  & $0.380\pm0.008$  & $5920\pm190$    & $1.012\pm0.007$     &$3.2\times10^{-5}\pm5\times10^{-6}$      & 1/6\\
                 & 40--60\% &$0.130\pm0.005$  & $0.350\pm0.012$  & $5650\pm230$    & $1.004\pm0.008$     &$7.3\times10^{-6}\pm7\times10^{-7}$      & 2/6\\
                 & 60--80\% &$0.126\pm0.007$  & $0.295\pm0.011$  & $5500\pm200$    & $1.001\pm0.004$     &$2\times10^{-7}\pm5\times10^{-8}$        & 6.5/2\\
\cline{2-8}
\end{tabular}%
\end{center}}
\end{table*}
\begin{table*}
{\scriptsize Table 1. Continue. \vspace{-.50cm}
\begin{center}
\begin{tabular}{ccccccccccc}\\ \cline{2-8}\cline{2-8}
Au-Au         & 0--10\%   &$0.150\pm0.006$  & $0.430\pm0.009$  & $7100\pm180$    & $1.01\pm0.006$      &$4.5\times10^{-4}\pm4\times10^{-5}$      & 2/7\\
  200 GeV       & 10--20\%  &$0.145\pm0.006$  & $0.414\pm0.012$  & $6900\pm200$    & $1.018\pm0.008$     &$1.7\times10^{-4}\pm3\times10^{-5}$      & 4/7\\
                 & 20--40\% &$0.140\pm0.007$  & $0.401\pm0.012$  & $6730\pm175$    & $1.013\pm0.008$     &$5.8\times10^{-5}\pm6\times10^{-6} $     & 1.5/7\\
                 & 40--60\% &$0.136\pm0.006$  & $0.368\pm0.010$  & $6600\pm190$    & $1.011\pm0.007$    &$1.3\times10^{-5}\pm4\times10^{-6}$       & 1.3/7\\
                 & 60--80\% &$0.131\pm0.007$  & $0.334\pm0.013$  & $6400\pm213$    & $1.001\pm0.006$    &$1.8\times10^{-6}\pm5\times10^{-7}$       & 2/6\\
\hline
Fig. 3           & 0--10\%      &$0.125\pm0.005$  & $0.319\pm0.010$   & $2600\pm155$   & $1.017\pm0.004$  &$2\times10^{-4}\pm4\times10^{-5}$     & 5.7/3\\\
Au-Au            & 10--20\%     &$0.120\pm0.006$  & $0.309\pm0.009$  & $2400\pm160$    & $1.04\pm0.004$   &$8\times10^{-5}\pm5\times10^{-6}$     & 2/3\\
7.7 GeV          & 20--40\%     &$0.116\pm0.007$  & $0.270\pm0.011$  & $2200\pm140$    & $1.013\pm0.005$  &$2.2\times10^{-5}\pm4\times10^{-6}$   & 5/3\\
Triton           & 40--80\%     &$0.112\pm0.005$  & $0.235\pm0.009$  & $2080\pm157$    & $1.005\pm0.004$  &$4.5\times10^{-6}\pm5\times10^{-7}$   & 1/1\\
\cline{2-8}
 Au-Au           & 0--10\%     &$0.136\pm0.007$  & $0.339\pm0.010$  & $3500\pm150$     & $1.018\pm0.005$  &$4.2\times10^{-5}\pm3\times10^{-6}$    & 0.1/3\\
 11.5 GeV        & 10--20\%    &$0.131\pm0.005$  & $0.320\pm0.009$  & $3320\pm165$     & $1.013\pm0.002$  &$1.5\times10^{-5}\pm5\times10^{-6}$   & 4/3\\
                 & 20--40\%    &$0.127\pm0.007$  & $0.280\pm0.011$  & $3110\pm160$     & $1.014\pm0.003$   &$3.6\times10^{-6}\pm6\times10^{-7}$   & 3.6/3\\
                 & 40--80\%    &$0.123\pm0.006$  & $0.250\pm0.010$  & $2800\pm152$     & $1.007\pm0.003$  &$7.9\times10^{-7}\pm5\times10^{-8}$   & 1.2/2\\
\cline{2-8}
  Au-Au          & 0--10\%    &$0.145\pm0.006$  & $0.369\pm0.011$  & $4000\pm205$      & $1.003\pm0.003$    &$2.5\times10^{-5}\pm6\times10^{-6}$  & 2.3/3\\
  14.5 GeV      & 10--20\%    &$0.140\pm0.007$  & $0.339\pm0.010$  & $3800\pm200$      & $1.005\pm0.005$   &$7\times10^{-6}\pm5\times10^{-7}$    & 2.5/3\\
                 & 20--40\%   &$0.136\pm0.005$  & $0.300\pm0.009$  & $3600\pm180$      & $1.004\pm0.006$   &$1.9\times10^{-6}\pm5\times10^{-7}$  & 4.4/3\\
                 & 40--80\%   &$0.130\pm0.005$  & $0.269\pm0.008$  & $3400\pm170$      & $1.0002\pm0.0004$ &$9\times10^{-7}\pm7\times10^{-8}$    & 2.3/1\\
\cline{2-8}
  Au-Au          & 0--10\%    &$0.145\pm0.006$  & $0.369\pm0.008$  & $4000\pm210$     & $1.004\pm0.004$    &$1\times10^{-5}\pm2\times10^{-6}$     & 2/3\\
  19.6 GeV       & 10--20\%   &$0.140\pm0.005$  & $0.339\pm0.009$  & $3800\pm200$     & $1.01\pm0.002$    &$3.5\times10^{-6}\pm3\times10^{-7}$    & 4.2/3\\
                 & 20--40\%   &$0.136\pm0.007$  & $0.300\pm0.011$  & $3600\pm207$     & $1.01\pm0.005$    &$8\times10^{-7}\pm5\times10^{-8} $     & 3.5/3\\
                 & 40--80\%   &$0.130\pm0.005$  & $0.270\pm0.010$  & $3400\pm195$     & $1.007\pm0.004$    &$1.6\times10^{-7}\pm4\times10^{-8}$    & 0.5/2\\
\cline{2-8}
  Au-Au         & 0--10\%    &$0.145\pm0.006$  & $0.369\pm0.012$   & $4000\pm140$     & $1.001\pm0.006$    &$5.4\times10^{-6}\pm6\times10^{-7}$     & 2.3/3\\
  27 GeV        & 10--20\%   &$0.140\pm0.005$  & $0.339\pm0.010$   & $3800\pm160$     & $1.012\pm0.004$    &$1.6\times10^{-6}\pm5\times10^{-7}$     & 2.6/3\\
                 & 20--40\%  &$0.136\pm0.005$  & $0.300\pm0.010$   & $3600\pm115$     & $1.02\pm0.005$     &$5.3\times10^{-7}\pm4\times10^{-8}$    & 2.7/3\\
                 & 40--80\%  &$0.130\pm0.005$  & $0.270\pm0.008$   & $3400\pm120$     & $1.007\pm0.004$    &$1\times10^{-7}\pm3\times10^{-8}$      & 1/2\\
\cline{2-8}
  Au-Au         & 0--10\%   &$0.145\pm0.004$  & $0.369\pm0.008$  & $4000\pm126$     & $1.01\pm0.006$      &$2.5\times10^{-6}\pm5\times10^{-7}$     & 5.6/3\\
  39 GeV        & 10--20\%  &$0.140\pm0.006$  & $0.339\pm0.012$  & $3800\pm143$     & $1.01\pm0.005$      &$1.1\times10^{-6}\pm3\times10^{-7}$     & 2/3\\
                & 20--40\%  &$0.136\pm0.007$ & $0.300\pm0.010$   & $3600\pm171$     & $1.02\pm0.003$      &$2.9\times10^{-7}\pm6\times10^{-8}$     & 0.5/3\\
                & 40--80\%  &$0.130\pm0.005$ & $0.270\pm0.011$   & $3400\pm165$     & $1.01\pm0.005$     &$7\times10^{-8}\pm4\times10^{-9}$       & 1.5/3\\
\cline{2-8}
  Au-Au         & 0--10\%  &$0.158\pm0.005$   & $0.388\pm0.009$  & $5000\pm170$    & $1.005\pm0.003$      &$1.26\times10^{-6}\pm3\times10^{-7}$      & 4/1\\
  62.4 GeV       & 10--20\% &$0.154\pm0.006$  & $0.359\pm0.010$  & $4800\pm180$    & $1.01\pm0.004$      &$4.5\times10^{-7}\pm5\times10^{-8}$       & 2/1\\
                 & 20--40\% &$0.150\pm0.007$  & $0.328\pm0.012$  & $4600\pm200$    & $1.025\pm0.002$     &$1.5\times10^{-7}\pm6\times10^{-8}$       & 0.7/1\\
                 & 40--80\% &$0.145\pm0.007$  & $0.300\pm0.009$  & $4415\pm175$    & $1.017\pm0.003$     &$3\times10^{-8}\pm5\times10^{-9}$         & 2/1\\
\cline{2-8}
  Au-Au         & 0--10\%   &$0.166\pm0.006$  & $0.400\pm0.012$  & $5750\pm180$    & $1.001\pm0.004$    &$4.2\times10^{-7}\pm4\times10^{-8}$       & 2/0\\
  200 GeV       & 10--20\%  &$0.162\pm0.006$  & $0.375\pm0.011$  & $5570\pm205$    & $1.01\pm0.0003$    &$2\times10^{-7}\pm4\times10^{-8}$         & 2.6/0\\
                & 20--40\% &$0.157\pm0.007$  & $0.342\pm0.009$   & $5300\pm190$    & $1.001\pm0.004$    &$7\times10^{-8}\pm5\times10^{-9} $        & 4.2/0\\
                & 40--80\% &$0.152\pm0.006$  & $0.310\pm0.009$   & $5111\pm215$    & $1.01\pm0.004$     &$1.75\times10^{-8}\pm3\times10^{-9}$      & 5/0\\
\hline
\hline
\end{tabular}%
\end{center}}
\end{table*}
\begin{figure*}[htbp]
\begin{center}
\includegraphics[width=14.cm]{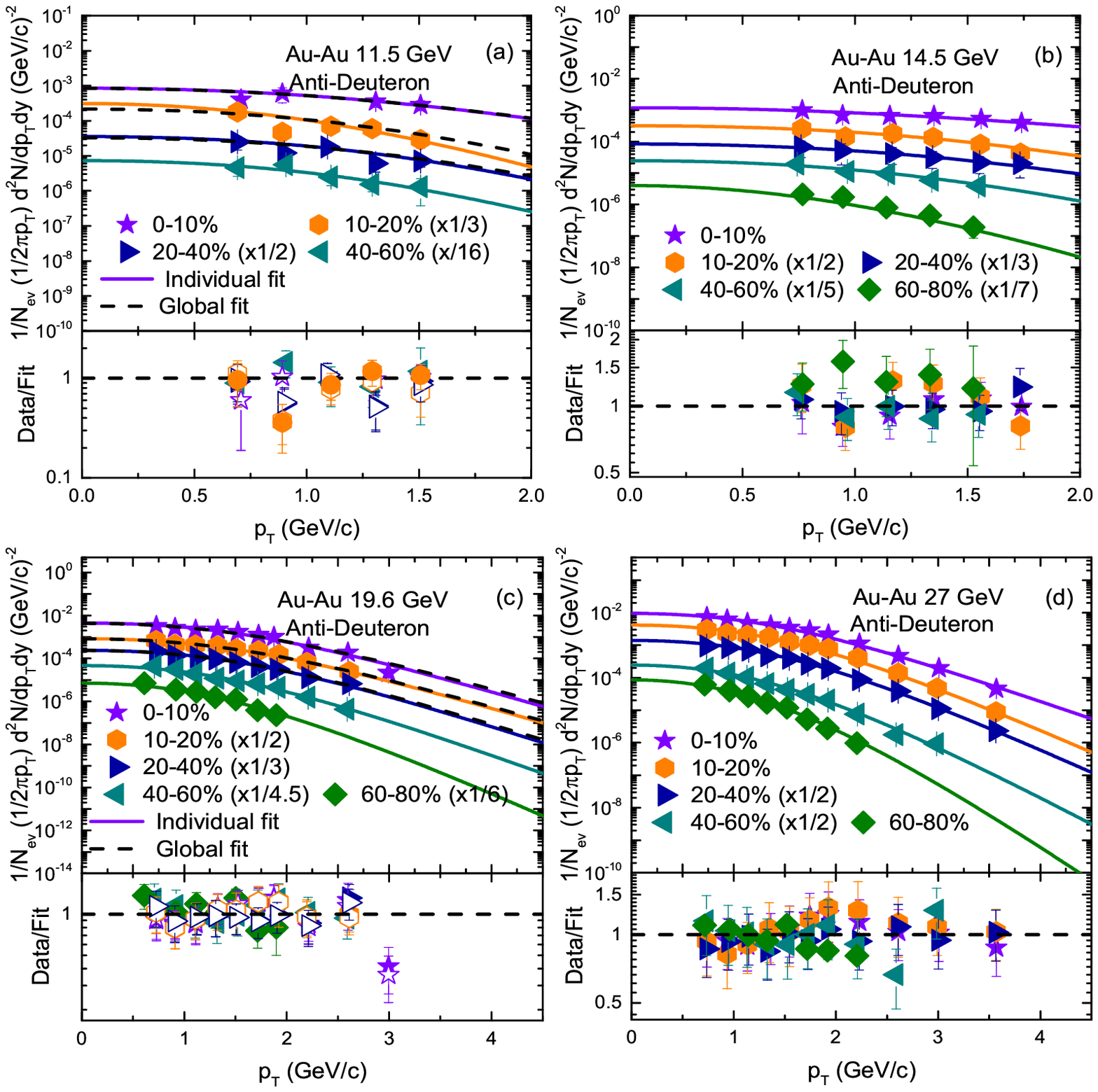}
\end{center}
continue
\end{figure*}
\begin{figure*}[htbp]
\begin{center}
\includegraphics[width=14.cm]{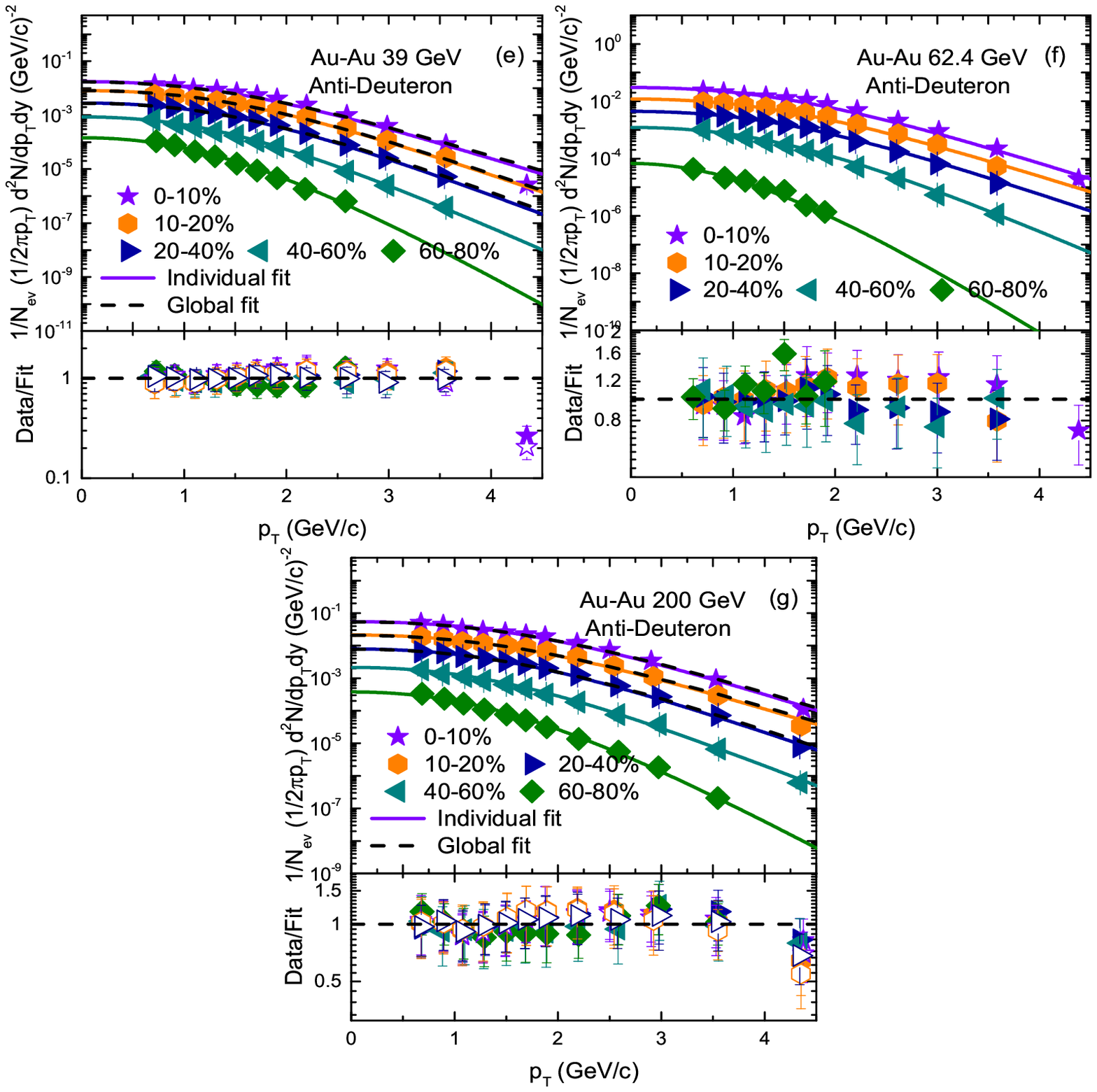}
\end{center}
Fig.2. Transverse momentum spectra of anti-deuteron produced at different energies in different centrality intervals in Au-Au, where $1/N_{ev}$ on the vertical axis denote the number of events. The symbols demonstrate the experimental data of STAR Collaboration measured in mid-rapidity $|y|<0.3$ \cite{59}. The curves are our results calculated by using the blast wave model with Tsallis statistics. The solid and dashed curves represents the individual and global fitting respectively.
\end{figure*}

Figure 2 is similar to figure 1, but it shows the $p_T$ spectra of anti-deuteron in Au-Au collisions at various energies in various centrality classes. The symbols are the experimental data of STAR Collaboration \cite{59} at mid-rapidity $|y|<0.3$  and the curves are our result by using the blast wave model with Tsallis statistics. The solid curves represent the individual fit, and the dashed curve (in some spectra) represent the global fit, while the results of their ratio of data/fit are denoted by filled and empty symbols respectively. The results of data/fit for the individual (global) fit are presented in the panels to monitor the departure of the individual (global) fit from data. One can see that Eq. (1) fits well the experimental data at 11.5, 14.5, 19.6, 27, 39, 62.4 and 200 GeV in Au-Au collisions.

Figure 3 is similar to figure 1 and 2, but it exhibits the $p_T$ spectra of triton in Au-Au collisions at different energies in different centrality classes. The symbols shows the experimental data of STAR Collaboration \cite{60} at mid-rapidity $|y|<0.5$  and the curves are the result of our fitting by the blast wave model with Tsallis statistics. The solid curves represent the individual fit, and the dashed curve (in some spectra) represent the global fit, while the results of their ratio of data/fit are denoted by filled and empty symbols respectively. The results of data/fit for the individual (global) fit are presented in the panels to monitor the departure of the individual (global) fit from data. One can see that Eq. (1) fits well the experimental data at 7.7, 11.5, 14.5, 19.6, 27, 39, 62.4 and 200 GeV in Au-Au collisions.
\begin{figure*}[htbp]
\begin{center}
\includegraphics[width=14.cm]{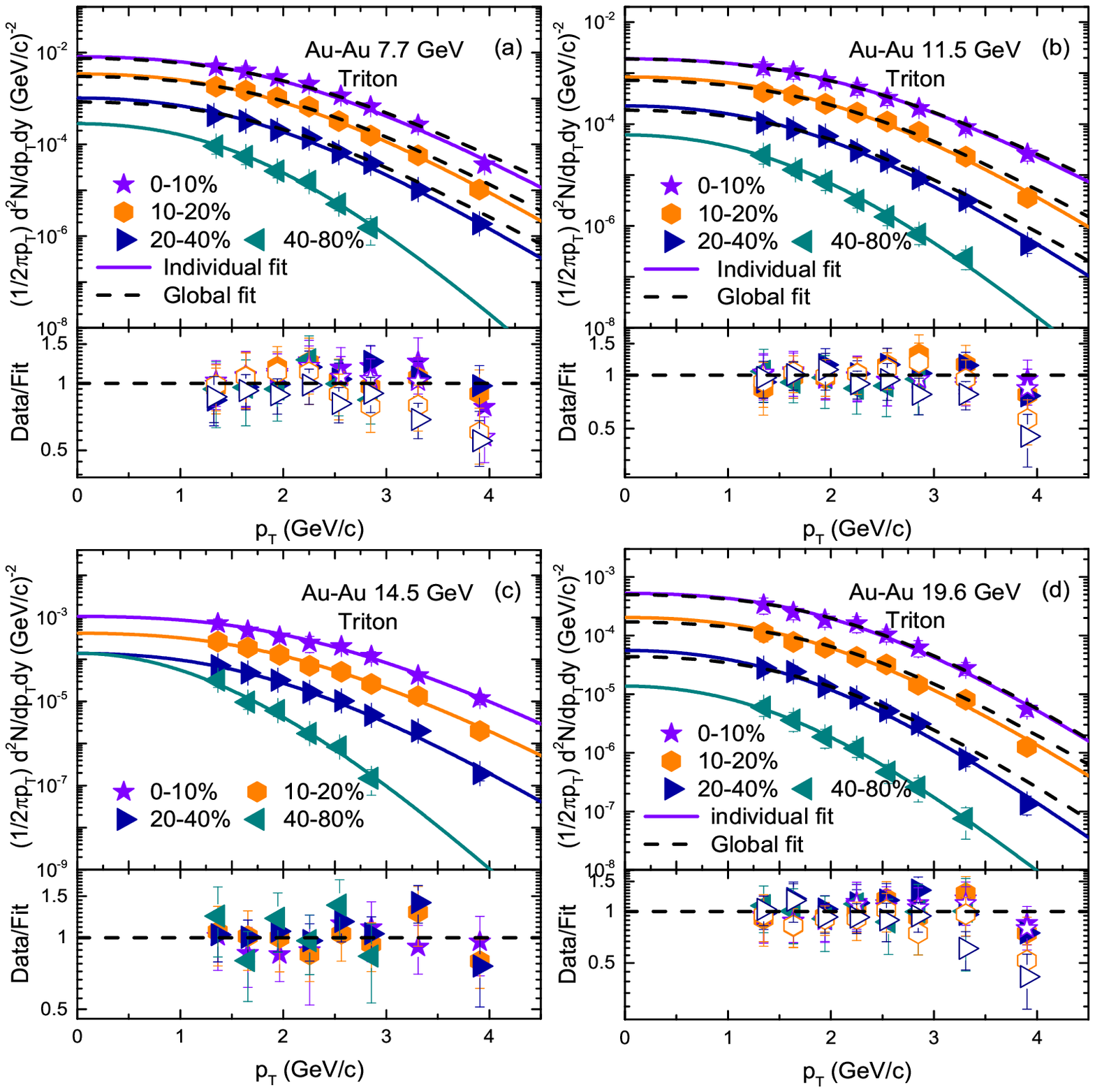}
\end{center}
continue
\end{figure*}
\begin{figure*}[htbp]
\begin{center}
\includegraphics[width=14.cm]{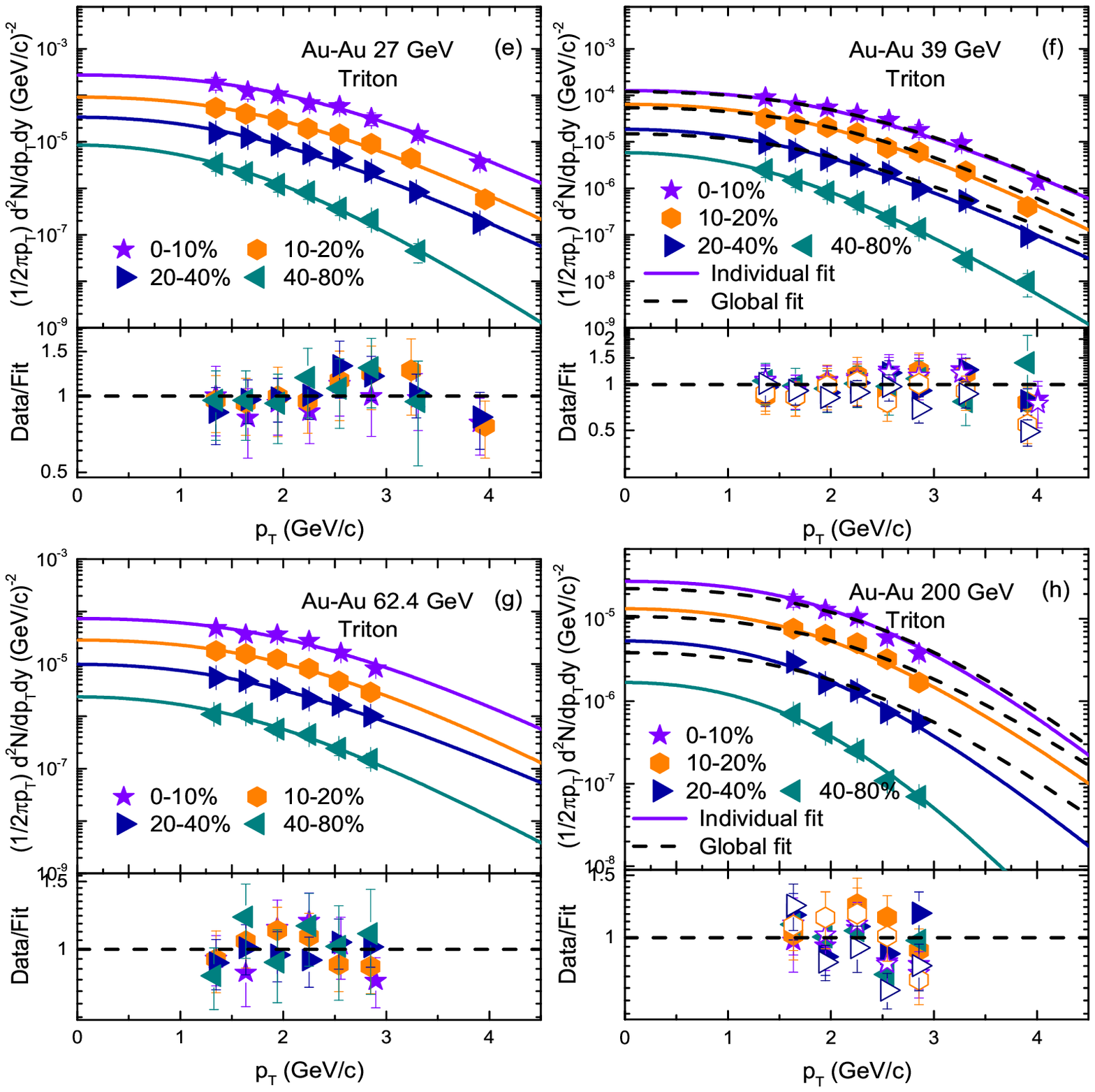}
\end{center}
Fig.3. Same as fig.1, but it shows the spectra of triton produced at different energies in different centrality intervals in Au-A collisions at $|y|<0.5$. The experimental data of STAR Collaboration are taken from the ref. 
cite{60}. The solid and dashed curves represents the individual and global fitting respectively.
\end{figure*}

 Figure 4 and figure 5 exhibits the dependence of $T_0$ and $\beta_T$ on energy and centrality. In figure 4, panel (a), (b) and (c) show the results of $T_0$ extracted from the individual fit for deuteron, anti-deuteron and triton respectively, and panel (d) shows $T_0$ extracted from the global fit. However, figure 5 (a), (b) and (c) shows the results of $\beta_T$ for deuteron, anti-deuteron and triton respectively, and panel (d) shows the results extracted for $T_0$ from the global fitting. In figure 4 and 5 the symbols from up to downward shows the centrality dependence of $T_0$ and $\beta_T$ respectively, while from left to right is the energy dependence of $T_0$ and $\beta_T$ respectively. One can see a saturation in $T_0$ and $\beta_T$ from 14.5 GeV-39 GeV. There are three energy regions which have three different mechanisms e;g $\sqrt{s_{NN}}$$<$14.5 GeV, where the system is baryon influenced according to our belief, and the system in $\sqrt{s_{NN}}$$>$39 GeV is meson influenced. There is no phase transition from hadronic matter to QGP in baryon dominated system due to small deposition of energy, while in meson dominated system the phase transition from hadronic matter to QGP occurs. In the region 14.5-39 GeV, the phase transition from hadronic matter to QGP seems to be possibly start in the system from 14.5 GeV in part volume due to deposition of large energy that goes from baryon influenced to meson influenced due to larger and larger volume in $\sqrt{s_{NN}}$=14.5-39 GeV and this suggest that 7.7 GeV and 39 GeV seems to be the onset energy of the partial phase transition and the whole phase transition respectively. The critical energy range is seem to be possibly exist in the range of 14.5 GeV to 39 GeV.

In the above discussion if the plateau in the second region from 14.5 to 39 GeV is regarded in the excitation functions of $T_0$ and $\beta_T$ as a reflection of the formation of QGP liquid drop, then the quick increase of $T_0$ and $\beta_T$ in the third region (39-200 GeV) is a reflection of higher temperature QGP liquid drop due to larger energy deposition, and larger energy density and blast wave, and then higher $T_0$ and $\beta_T$ should be created due to the higher collision energy.
Though any temperature needs to be bound by the phase transition on one hand and free streaming on the other hand, larger energy deposition at higher energies may heat the system to a higher temperature. It is also possible to heat the formed QGP and hadronized products to higher temperature.

In some cases, the quality of fits is not sufficient, our main conclusion that the rise of temperature below 14.5 GeV is the indication of de-confinement of hadronic matter to QGP is weak. The information of phase transition happened at higher temperatures and near the chemical freezeout may be reflected at the kinetic freeze-out of a hadronic system. The plateau structure in the second region in the excitation function $T_0$ is expected to relate to the phase transition, but this relation is not clear at present, and other works related to this issue are needed to make a strong conclusion. In other words, we can say that it is a loose interpretation to conclude the onset of de-confinement just from the quality of some fits. More investigations and also comparisons with other findings are needed but this issue is beyond the scope of this analysis.

 We found that the kinetic freeze out temperature ($T_0$) is larger in central collisions than in peripheral collision and  this is in consistent with \cite{40,41,42,44,45,46}, but disagrees with \cite{36,43}. Furthermore, it is observed that triton freezeout earlier than deuteron and anti-deuteron and this result is consistent with \cite{19a,20a}, and has less values of $\beta_T$ than deuteron and-anti-deuteron which exhibits the mass dependence of $T_0$ and $\beta_T$. The lower values of $\beta_T$ for triton is due to the fact that triton can be left behind in the system evolution process due to their large masses. This maybe the possible reflection of hydrodynamic behavior \cite{61}, in which heavy particles are left due to their small velocity. However deuteron and anti-deuteron has same $\beta_T$. In central collision, the reaction is more violent due to large number of nucleons taking part in the reaction which experience a strong squeeze that results in a rapid expansion of the system and therefore $\beta_T$ is larger. However this reaction becomes less violent as we go towards periphery and the squeeze becomes weak due to the accumulation of less participant nucleons in the interaction that results in the steady expansion of the system and the values of $\beta_T$ become smaller.
\begin{figure*}[htbp]
\begin{center}
\includegraphics[width=14.cm]{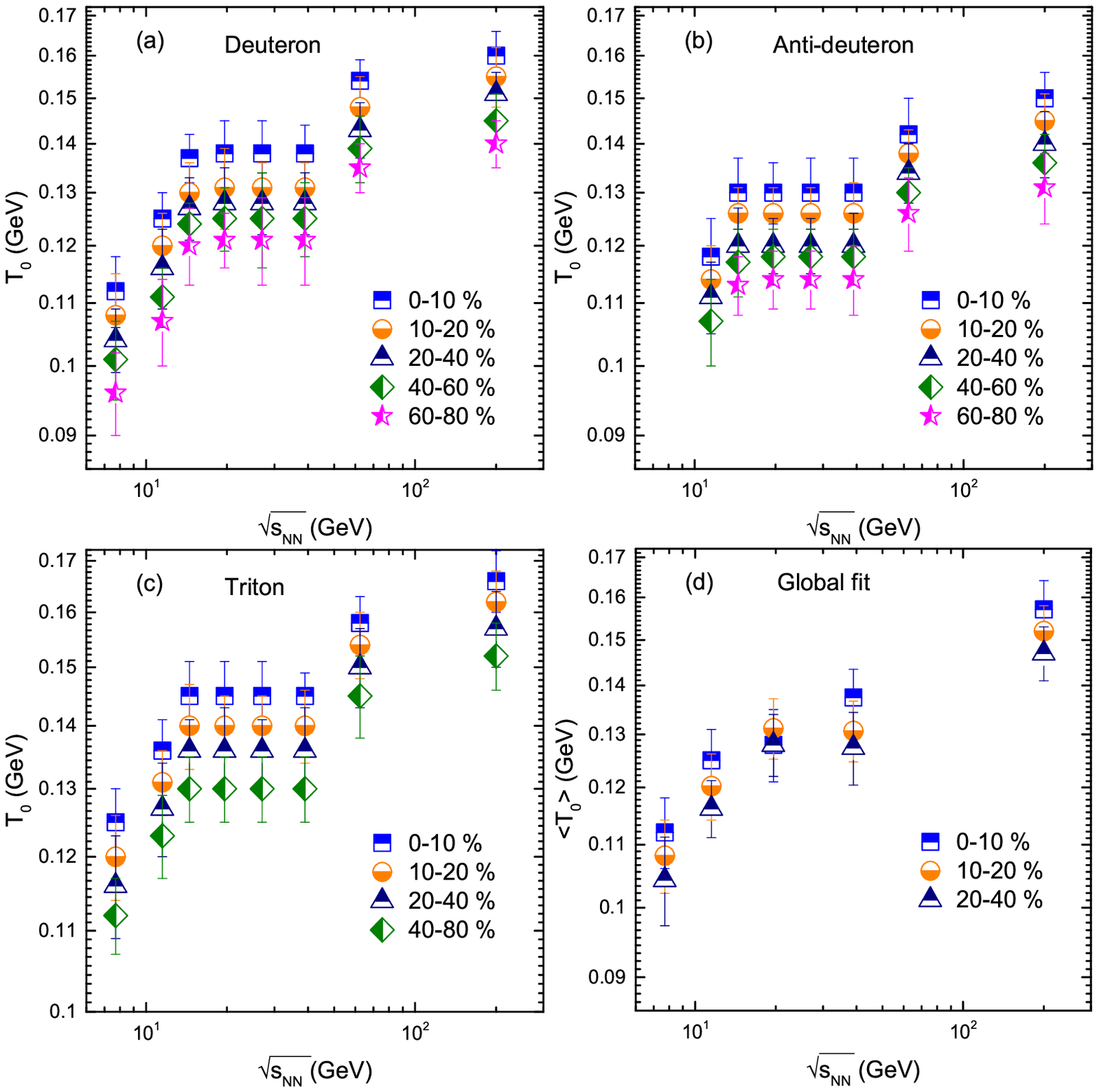}
\end{center}
Fig.4. $T_0$ dependence on energy and centrality for deuteron, anti-deuteron and triton in Au-Au collisions. Different symbols represent different energy and centrality bins.
\end{figure*}
\begin{figure*}[htbp]
\begin{center}
\includegraphics[width=14.cm]{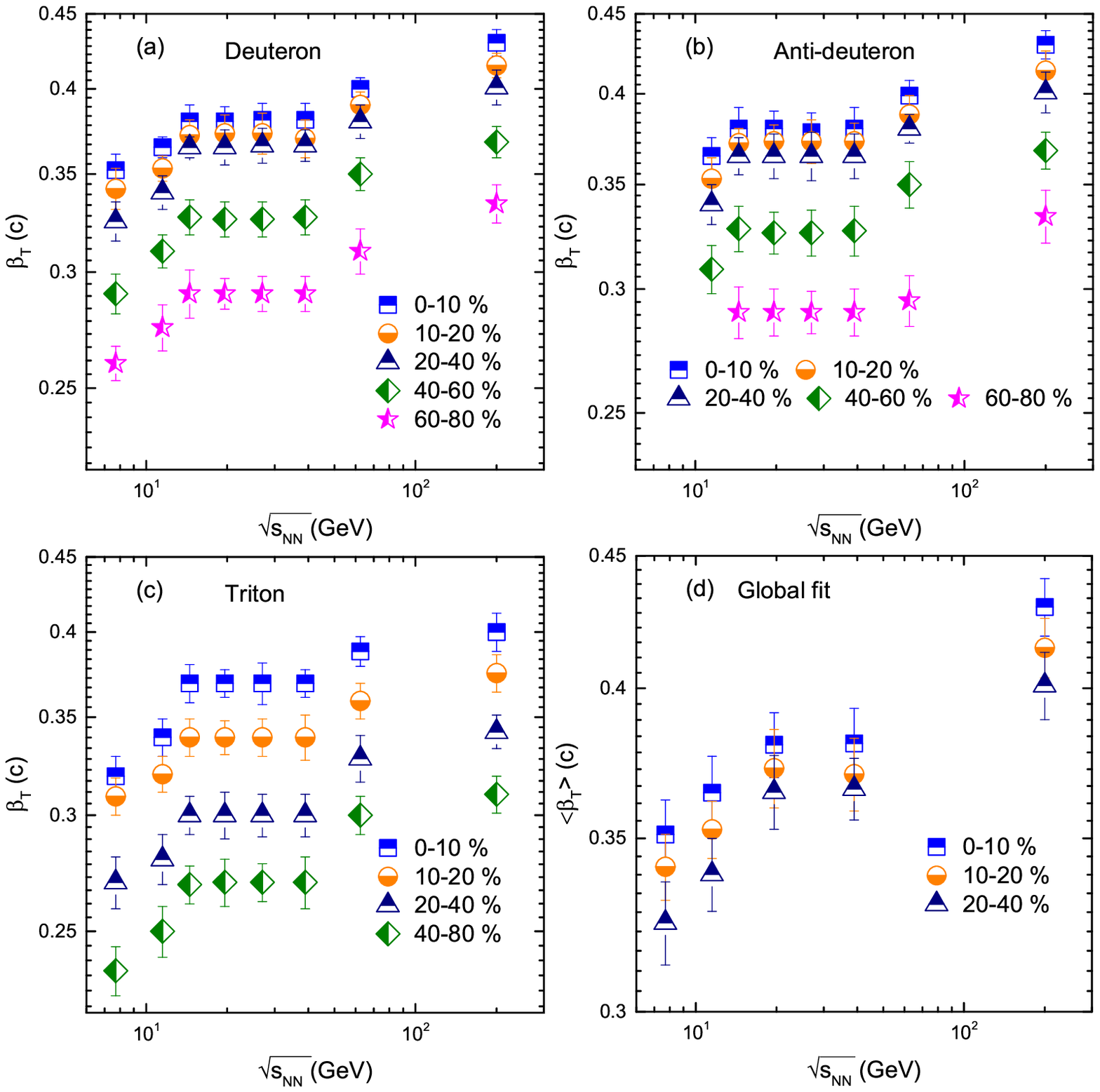}
\end{center}
Fig.5. $\beta_T$ dependence on energy and centrality for deuteron, anti-deuteron and triton in Au-Au collisions. Different symbols represent different energy and centrality bins.
\end{figure*}
\begin{figure*}[htbp]
\begin{center}
\includegraphics[width=14.cm]{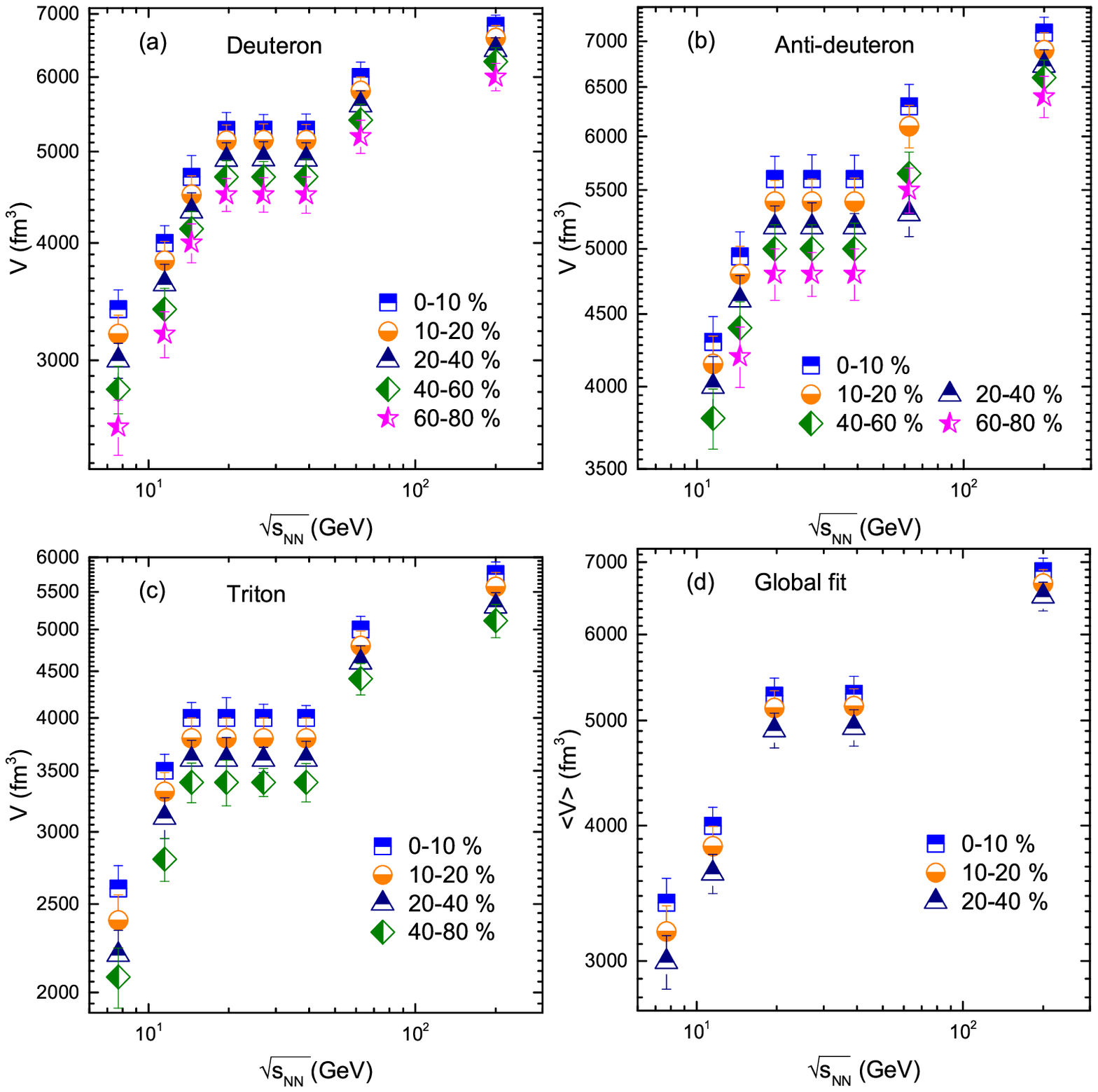}
\end{center}
Fig.6. Dependence $V$ on energy and centrality for deuteron, anti-deuteron and triton in Au-Au collisions. Different symbols represent different energy and centrality bins.
\end{figure*}
Figure 6 is similar to fig 4 and 4, but demonstrates the behavior of $V$ on energy and centrality. $V$ increase from 7.7 GeV to 19.6 GeV due to the fact that at higher energies, the evolution time of the system is longer, which corresponds to larger partonic system. At 19.6 GeV to 39 GeV, $V$ saturates, we believe it maybe due to phase transition. It is also observed that $V$ decrease from central to peripheral collisions due to large number of binary collisions in central collisions compared to the peripheral ones, because of re-scattering of partons that leads a quick approach of the system to equilibrium state.

It should be noted that light nuclei are formed by the nucleons coalescence with similar momenta. In present work, it is can be seen that although the mass of deuteron and anti-deuteron is the same but the value of kinetic freezeout temperature is larger for deuteron than anti-deuteron. We believe that this maybe due to the large coalescence of nucleons for deuteron than for the anti-deuteron, but it is hard to conclude that deuteron freezeout early because $d$ and $\bar d$ have nearly the same $V$ in error bars. In addition, $\beta_T$ is larger in central collision compared to peripheral collision.

 We would like to point out that although we fit the spectra of spectra of deuteron and triton individually
in Figs. 1, 2 and 3 by the solid curves, the average parameters weighted by particle yields $N_0$ in Table 1
should be used for the particles simultaneously by the dashed curves. Figures 4, 5 and 6 show that the average
parameters weighted by particle yields are closer to those for deuteron due to the fact that the yield of deuteron is the
most at the considered energy.
\begin{figure*}[htbp]
\begin{center}
\includegraphics[width=14.cm]{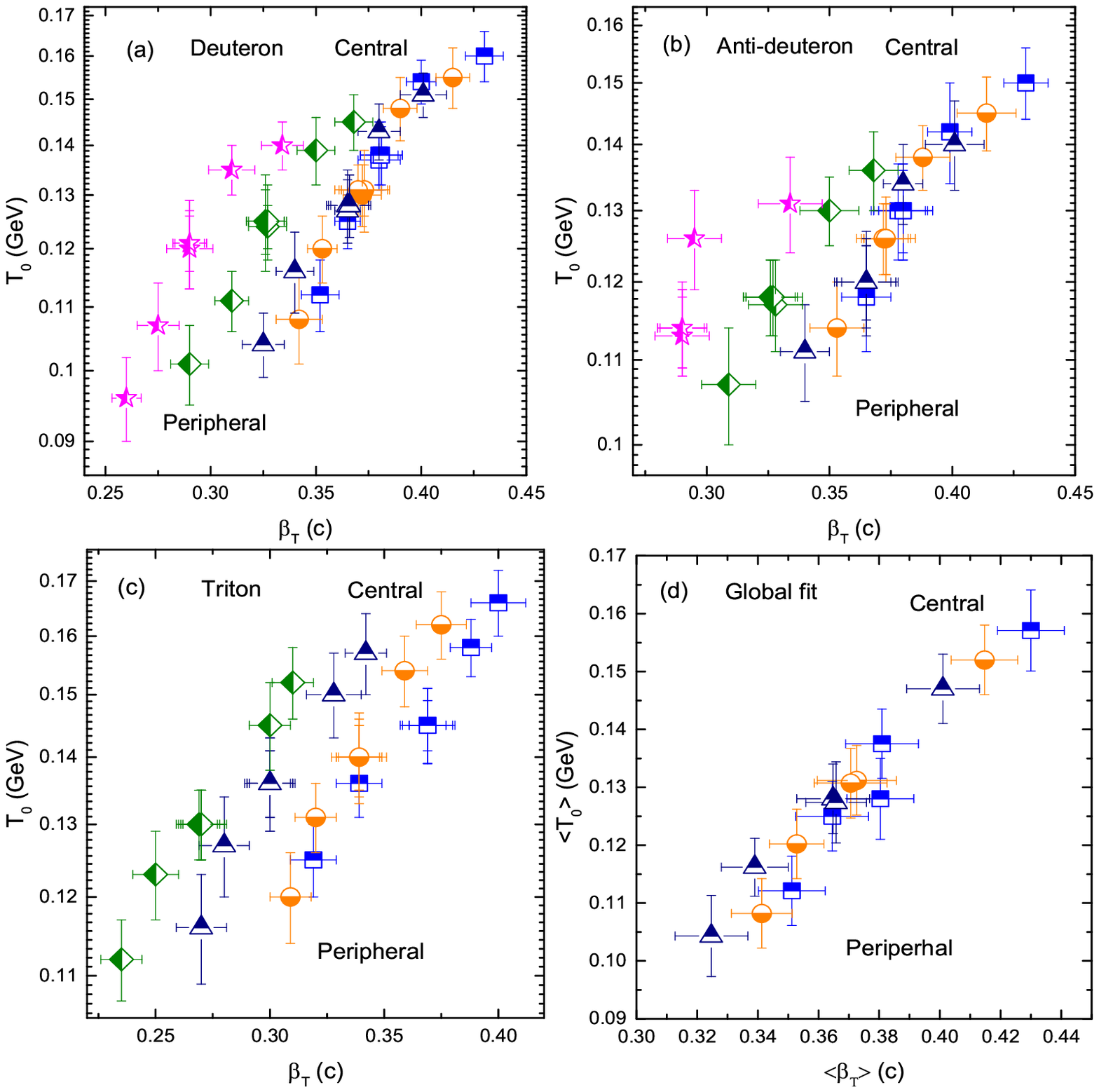}
\end{center}
Fig.7. Correlation of $T_0$ and $\beta_T$ in individual fit for deuteron in panel (a), anti-deuteron in panel (b) and triton in panel (c) are presented. While panel (d) shows the correlation of $T_0$ and $\beta_T$ in global fit.
\end{figure*}

 In addition, the variation of $T_0$ on $\beta_T$ for various collision energies and event centralities are presented in Fig. 7, where the symbols represent the parameter values averaged by weighting the yields of different particles. Panels (a), (b) and (c) show the correlation of $T_0$ and $\beta_T$ obtained from the individual fit, while panel (d) shows their correlation obtained from the global fit. It can be seen that $T_0$ increases with the increase of $\beta_T$ . At higher energy and in central collisions, larger $T_0$ and $\beta_T$ can be seen. There is a positive correlation between $T_0$ and $\beta_T$.

 The reason behind the increasing trend of $T_0$ and $\beta_T$ with the increase of collision energy is that more energies are deposited per particle in collisions at higher energies in the considered RHIC-BES energy range. Meanwhile, due to relativistic constriction effect, the system size at higher energy decreases, which results in a smaller volume, then a larger energy density and larger $T_0$. Meanwhile, at higher energy, the squeeze is more violent, which results in a rapider expansion and larger $\beta_T$.

The central collisions contain more nucleons than the peripheral collisions, then more energies per particle are deposited in central collisions and the temperature is high in central collisions than in the peripheral collisions. Meanwhile a rapid expansion appears due to more violent squeeze in central collisions, comparatively to peripheral collisions. The fireball is hot in central collision and it expands rapidly, but if the collisions become peripheral, the temperature of the fireball becomes lower and expansion becomes steady. In some literature, it is opposite, where there is smaller $T_0$ and larger $\beta_T$ in central collisions and the correlation is negative. This difference is caused by different model and different methods. Both the positive and negative correlations have different approaches of understanding. The negative correlation indicates that the thermal motion ($T_0$) is transformed into collective motion ($\beta_T$) as the system cools down. While the interpretation of positive correlation in the present work is that, in central collision more energy per particle is stored than in peripheral collisions, which leads to a faster expansion, and, hence, to both a larger flow velocity and a higher kinetic freezeout temperature.
\begin{figure*}[htbp]
\begin{center}
\includegraphics[width=14.cm]{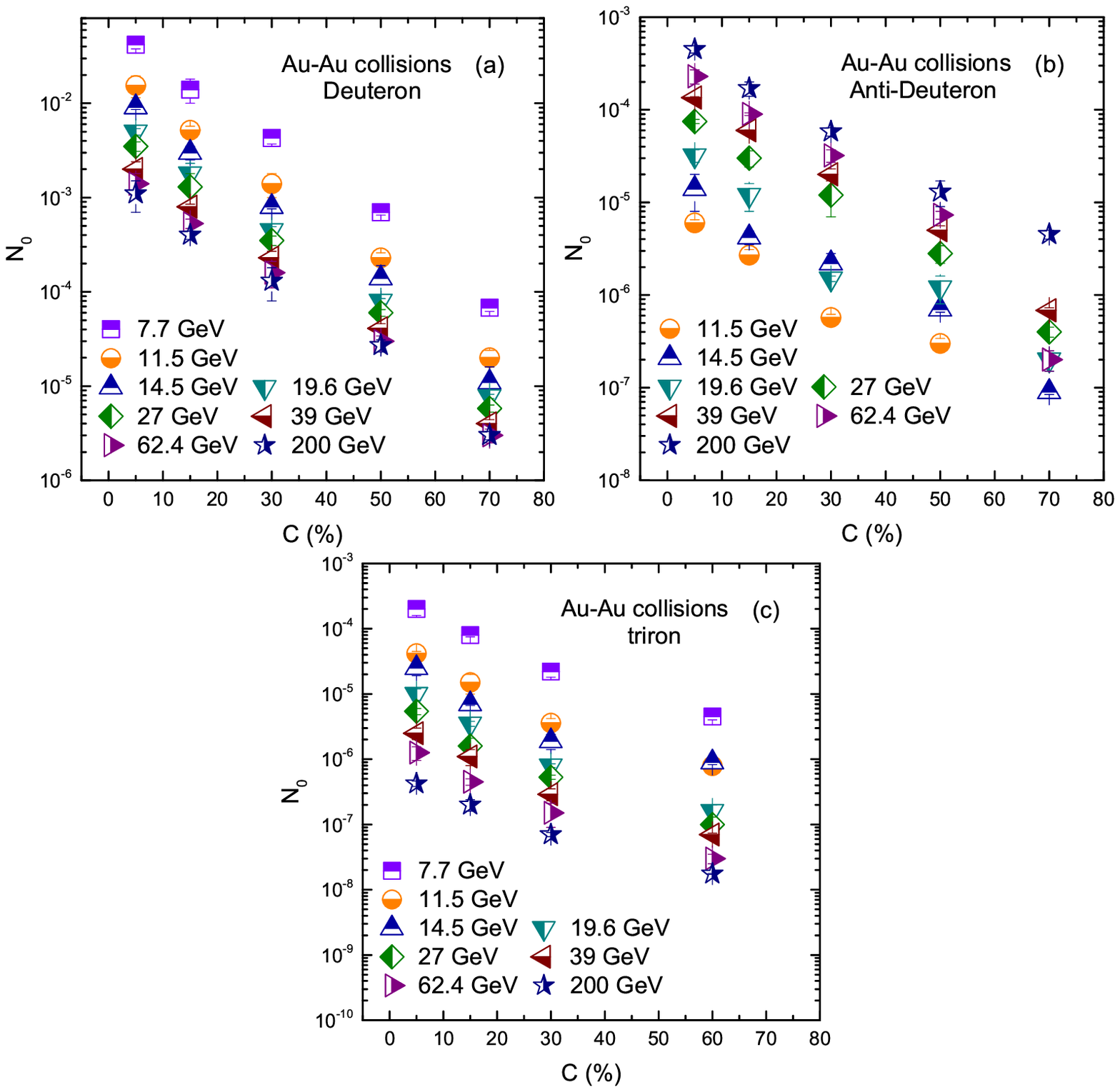}
\end{center}
Fig.8. Dependence $N_0$ on energy and centrality for deuteron, anti-deuteron and triton in Au-Au collisions. Different symbols represent different energy and centrality bins.
\end{figure*}

Figure 8 shows the dependence on $N_0$ on energy and centrality. The symbols from up to downward shows the energy dependence of $N_0$ while from left to right it shows the centrality dependence. $N_0$ increase with increasing energy and decrease from central to peripheral collisions.

It should be noted that that we analyze the global fitting in a few cases in order to show that the trend of the parameters ($T_0$ and $\beta_T$ is the same, only the concrete values are different.)

 Before going to conclusions, it is very important to point out that the values of $q$ are shown in table 1. As an entropy index, $q$ characterizes the degree of equilibrium of the system. In general, an equilibrium state corresponds to $q$=1. The values of q obtained in this work are approximately close to 1,which indicates that the system in the considered centralities and energy range stays approximately in an equilibrium state or in a few local equilibrium states. This also exhibits that the blast-wave fit approximately useable in this work.

However, it is noteworthy that the entropy index $q$ is a very sensitive quantity. A large $q$ which is not close to 1 results in a wide distribution, and a small $q$ which is close to 1 results in a narrow distribution. In the present work the values of $q$ in some centrality intervals and some energies are very close to 1 (1.0001) , while it is a fact that $q$ = 1.01 is not close enough to 1. Therefore, it does not imply that the Tsallis blast-wave fit is close to its Boltzmann Gibbs counterpart if we use the same $T_0$ and $\beta_T$ in the case of $q$ = 1.01. To reduce the difference between the Tsallis blast-wave fit and its Boltzmann Gibbs counterpart, we need
$q$ = 1.0001 or the one which is closer to 1 in all centrality bins at all energies.
\\
 {\section{Conclusions}}
 The main observations and conclusions are summarized here.

 a) The transverse momentum spectra of deuteron, anti-deuteron and triton are analyzed by the blast wave model with Tsallis statistics and the bulk properties in terms of the kinetic freezeout temperature, transverse flow velocity and kinetic freezeout volume are extracted.

 b) The kinetic freezeout temperature ($T_0$) and transverse flow velocity ($\beta_T$) are observed to follow an increasing trend below 14.5 GeV. However it remains constant from 14.5-39 GeV, it may indicate that phase transition in part volume starts at 14.5 GeV and ends at 39 GeV, and the critical point is supposed to be in the range of 14.5-39 GeV, while the kinetic freezeout volume ($V$) increases with energy due to longer evolution time at higher energies and saturates from 19.6 GeV to 39 GeV.

 c) We observed that $T_0$ decrease from central to peripheral collisions. while $\beta_T$ decrease from from central to peripheral collisions due to decreasing of number of participants in peripheral collisions and the collision becomes less violent and the system does not expand so rapidly.

 d) $V$ decrease from central to peripheral collisions due to the fact that the number of binary collisions due to re-scattering of partons with each other become less in peripheral collisions.

 f) Triton freezeout earlier than deuteron and anti-deuteron due to its larger mass. In addition, triton has less values of $\beta_T$ than deuteron anti-deuteron because triton are left behind in the system evolution process due to its large mass.

 g) The parameter $q$, which characterizes the degree of equilibrium of the system, in some cases are observed to be close to 1, which renders that the system in the considered centralities and energy range stays approximately in an equilibrium state or in a few local equilibrium states. This also indicates that the blast-wave fit approximately useable in this work.

)
\\
\\

{\bf Data availability}

The data used to support the findings of this study are included
within the article and are cited at relevant places within the
text as references.
\\
\\
{\bf Compliance with Ethical Standards}

The authors declare that they are in compliance with ethical
standards regarding the content of this paper.
\\
\\
{\bf Acknowledgements}

The authors would like to thank support from the National Natural Science
Foundation of China (Grant Nos. 11875052, 11575190, and 11135011), the National Natural Science Foundation of China (Grant No. 12175115), the
Natural Science Foundation of Shandong Province, China (Grant No. ZR2020MA097). We would also would like to acknowledge the support of Ajman University Internal Research Grant NO. [DGSR Ref.2021-IRG-HBS-12].
\\
\\

\end{multicols}
\end{document}